\documentclass[12pt,preprint]{aastex}
\pdfoutput=1
\usepackage{graphicx}
\usepackage{epsf}
\usepackage{natbib}
\def\GC{Galactic center}
\def\sgr{Sgr~A$\ast$}
\def\Msolar{M$_\odot$}
\def\figwidth{5.0in}
\def\erf{{\rm erf}}

\begin{document}

\title{Gravitationally Focused Dark Matter Around Compact Stars}

\author{Benjamin C. Bromley}
\affil{Department of Physics \& Astronomy, University of Utah, 
\\ 115 S 1400 E, Rm 201, Salt Lake City, UT 84112}
\email{bromley@physics.utah.edu}

\begin{abstract}

If dark matter self-annihilates then it may produce an observable
signal when its density is high. The details depend on the intrinsic
properties of dark matter and how it clusters in space. For example,
the density profile of some dark matter candidates may rise steeply
enough toward the Galactic Center that self-annihilation may produce
detectable $\gamma$-ray emission.  Here, we discuss the possibility
that an annihilation signal may arise near a compact object (e.g., neutron
star or black hole) even when the density of dark matter in the
neighborhood of the object is uniform. Gravitational focusing produces
a local enhancement of density, with a profile that falls off
approximately as the inverse square-root of distance from the compact
star. While geometric dilution may overwhelm the annihilation signal
from this local enhancement, magnetic fields tied to the compact
object can increase the signal's contrast relative to the background.

\end{abstract}

\section{Introduction}

Dark matter accounts for the majority of the mass in the Universe, yet
its identity remains elusive. Candidates include weakly interacting
massive particles (WIMPs) like the neutralino ($\chi$), the
supersymmetric partner of the neutrino~\citep{PagPri82}, although
their properties are only loosely constrained by theory and
experiment.  In some cases, plausible values of the mass and cross
section suggest that self-annihilation signatures may be detectable in
regions where the density of dark matter is
high~\citep{BerGurZyb92,BerGon96,BerHooSil04}.  For example,
\citet{BerUllBuc98} calculate the gamma ray flux from neutralino
self-annihilation in the \GC\ \citep[see also][]{ZahHoo06}, while
\citet{Tyl02} and \citet{BerHoo06} provide estimates of the
annihilation signal from the nearby Draco dwarf
galaxy. Intermediate-mass black holes may yield a WIMP annihilation
signal \citep{BerZenSil05}, as may remnant dark matter minihalos
distributed throughout the Galaxy \citep[e.g.][]{BerDokEro03,SanDieFre11}.

The observability of an annihilation signal critically depends on the
number density of particles, since the local event rate is
proportional to the density squared. Estimates of the strength of this
signal typically derive from the assumption that dark matter density
profiles follow a power law, $\rho \sim r^{-\gamma}$ in many
astrophysical contexts. The power-law index $\gamma$ is between 1 and
2 in the central regions of galaxies according to cosmological
simulations \citep{NavFreWhi96, Mooetal99,Powetal03}, corresponding to
a density ``cusp.''  If the simulations realistically describe the
distribution of dark matter, then self-annihilation of WIMPS may
indeed be observable in the centers of galaxies.

The presence of massive black holes in the centers of galaxies may
further enhance the steep rise of a dark matter density
profile. \citet{GonSil99} model the adiabatic growth of a central
black hole to show that a density ``spike,'' with a profile steeper
than $r^{-2}$, forms around the black hole.  Dynamical processes, such
as scattering by stars and capture the black hole, may erode a spike
over time, although it may remain largely intact \citep{BerMer05}.
More problematic is the inspiral of smaller massive black holes
captured from accreted galaxies that may disrupt a density spike
\citep{Meretal02}. Binary black hole coalescence can destroy density
structures in the vicinity of the central black hole. However, even if
merger event disrupts a density spike, relaxation processes may
regenerate a dark matter ``crest,'' with a density profile that falls
off as $r^{-1.5}$ \citep{MerHarBer07}.  The time scale for the growth
of a dark matter crest can be long, $\sim$10~Gyr in the case of the
Milky Way, so that its structure could reflect the merger history of
the Galaxy.

From an observational perspective, strong density cusps or spikes are
not obviously common. For instance, \citet{Kraetal98} find that dwarf
and low surface brightness galaxies have shallow core profiles, with
$\gamma < 0.5$, consistent with results from \citet{vanSwa01} in a
study of dwarf galaxies.  \citet{Miletal02} identify an
anticorrelation between profile steepness and mass of the central
black hole, lending support for a scenario in which the black hole
grows through a sequence of merger events which tend to reduce or
destroy any density cusp.

Even if no density cusp or spike exists in a galactic nucleus, the
presence of a massive black hole, such as \sgr\ in the center of the
Milky Way \citep{MelFal01}, can facilitate annihilation radiation.
The reason is that gravitational focusing inevitably leads to a dark
matter density enhancement near the black hole.  This phenomenon will
occur for any star in a field of dark matter~\citep{DanCam57, Gri88,
SikWic02,AleGon06}.  The purpose of this paper is to review the gravitational
focusing effect and discuss its observational consequences in terms
of dark matter annihilation radiation.

\section{Gravitational focusing}

A point particle with mass $M$ in a uniform bath of hot,
dissipationless dark matter will influence dark matter orbits, but in a
manner which conserves phase-space density along each trajectory.
\citet{DanCam57} derived analytical expressions for the density
enhancement near the point mass in a bath of unbound particles with a
Maxwellian velocity distribution with velocity dispersion
$\sigma_v$. For a point mass which is at rest with respect to the bulk
dark matter, the density enchancement is
\begin{equation}
\label{eq:rho}
\frac{\rho}{\rho_\infty} =
\sqrt{\frac{2}{\pi}}\ q + e^{q^2/2}\left[1 - \erf(q/\sqrt{2})\right] ,
\end{equation}
where $q=\sqrt{2GM/\sigma_v^2r}$ is the ratio of escape velocity at
distance $r$ from the point mass.  Thus, close to the point mass the
density enhancement grows with decreasing radius according to
$r^{-0.5}$. Note that details of Dandy \&~Camm's results have been contested by
\citet{Gri88}, and more recently by \citet{SikWic02}. The final word
comes from \citet{AleGon06}, who resolve the discrepancies.

% I only report
% here that when gravitational focusing enhances the density by a factor
% of a few or more, as in the cases of interest in this work, all three
% results evidently agree.

To support the previous analytical work, and to provide a framework
for studying gravitational focusing about an arbitrary mass
distribution, I developed a computer code to track the flow of dark
matter about a compact star. The code calculates Monte Carlo orbits
assuming that the dark matter has a Maxwellian velocity distribution
at large distance from the star, and that the star may have some
finite speed relative to the average rest frame of the dark matter.
The resulting simulations show that the density enhancement from
gravitational focusing can exceed two orders of magnitude near the
surface of a neutron star, and even higher within a factor of a few
times the Schwarzschild radius of a black hole. Here, to account for
the possibility that a dark matter particle may be captured by the
black hole, the code does not track orbits inside of five times the
Schwarzschild radius \citep[see][]{GonSil99}.  The measure of particle
trajectories that reach inside this distance is small, and numerical
simulations demonstrate that these orbits do not greatly affect the
density enhancement at larger radii. Thus analytical calculations for
point particles are broadly applicable to compact objects.

Figure~1 illustrates the density enhancement at the \GC\ for a
background dark matter halo distribution similar to one calculated by
\citet{Meretal02}. The background model has a profile that falls off
as $r^{-1.0}$, as predicted by numerical simulations
\citep{NavFreWhi96}. A ``core'' has been imposed to crudely mimic the
effects of dark matter clearing by black hole mergers of
\citet{Meretal02}.  To calculate the effect of gravitational focusing
by \sgr, a black hole mass of $3\times 10^6$~\Msolar is assumed, and
the velocity distribution is assigned to be isotropic with a
dispersion of $\sigma_v=155$~km/s.  The black hole mass, the
assumption of an isotropic velocity distribution, and the value of
$\sigma_v$ used here are all approximately consistent with observed
stellar kinematics in the \GC~\citep{Gheetal98, Genetal00}.

The observed dark matter self-annihilation 
flux depend on the local emissivity,
\begin{equation}
\label{eq:j}
j = \frac{Y \left<\sigma_{ann}v\right> \rho_{\rm dm}^2}{4\pi m_{\rm dm}^2}
\end{equation}
where $\rho_{\rm dm}/m_{\rm dm}$ is the dark matter number density, and
$\left<\sigma_{ann} v\right>$ gives the self-annihilation rate per
unit density.  For the neutralino, typical values from
 the literature \citep[e.g.][]{BerGon96} are $m_{\chi}
\equiv m_{\rm dm} = 100$~GeV, and $\left<\sigma_{ann} v\right> =
10^{-26}$~cm$^3$/s, independent of pairwise closing speed $v$. The
quantity $Y$ specifies the yield of decay by-products; for example,
the bolometric yield corresponds to $Y=m_\chi c^2$. The emissivity per
unit frequency of photons produced by electrons in a magnetic field
depends on electron-positron production channels, as well as
synchrotron radiative efficiencies \citet[e.g.,][]{Tyl02}.

In the case of certain neutralino decay products, namely neutrinos,
the flux is a straightforward line-of-sight integral over the
emissivity, since self-absorption and diffusion do not
occur~\citep{GonSil99}. The line of sight integral along
some sky direction $\hat{n}$ is conveniently
expressed in dimensionless form as \citep{BerUllBuc98, Meretal02}
\begin{equation}
\label{eq:J}
J(\hat{n}) = 
\frac{1}{8.5\ {\rm kpc}}\left(\frac{1}{0.3\ {\rm GeV/cm}^3}\right)^2
\int_{\hat{n}} \, d\ell \, \rho_{\rm dm}^2 \, .
\end{equation}
Figure~2 gives $J$, averaged inside a circular aperture centered on
\sgr, as a function of aperture radius. The density profile is the
same as in Figure~1. The enhancement from gravitational focusing is
significant inside small apertures.  Even so, \citet{Beretal04} point
out that the neutrino flux from the Galactic Center will be
undetectable if the current gamma-ray constraints are any indication
of the annihilation rate. \citet{Erketal10} are more hopeful from a
theoretical perspective, while the observations are providing
limits to the neutrino flux \citep[e.g., from IceCube][]{Abbetal11},
but no Galactic Center signal at this point.

%just as it is
%in certain models with binary black hole mergers in the center of
%the Galaxy \citet{Meretal02}.

\section{Implications}

The strength of a gravitationally focused density profile around a
compact object is unfortunately insufficient to generally produce a
strong flux enhancement above that from the background density. In a
uniform bath of dark matter the flux only grows logarithmically with
decreasing radius, and is much weaker than in the case of the density
spike envisioned by \citet{GonSil99}.  Still, gravitational focusing
may have astrophysical relevance if the flux is enhanced by any
distinctive properties of the environment around the compact
object. This situation may occur in the case of synchrotron radiation
from a region around a black hole or neutron star.  For supermassive
black holes, magnetic field strengths can be orders of magnitude
higher than in the interstellar medium.  Just as the gravitational
focusing is pinned to the black hole, so is the magnetic field, as it
is likely tied to gas accreting on to the black hole.

%Even for \sgr, which is believed to have a low
%accretion rate \citep[see][and references therein]{MelFal01}, a
%magnetic field strength of the order of 10~G may exist near the black
%hole \citep{MelLiuCok01}. This value is roughly $10^6$ times larger than
%field strengths typical of the interstellar medium.

An estimate of the synchrotron flux from \sgr\ follows from the
density profile of Figure~1 and an equipartition model for the
magnetic field strength \citep{Mel92}, in which $B = (r/{\rm
pc})^{-1.2}$~mG.  The value for electron-positron yield may be obtained
as in \citet{Tyl02}; in the low energy limit, the electron energy
distribution per neutralino annihilation is $dN/dE \approx
E^{-1.5}$. An integration of the radiative transfer equation should include
the possible effects of synchrotron self-absorption. 
Synchrotron limits have been used by \citet{BerSigSil01} to place
constraints on dark matter properties. However, these limits depend on
a strong flux generated in a density cusp and spike, along with a
relatively strong magnetic field, all of which must be centered on 
\sgr.  According to the results of \citet{Meretal02},
even a relatively minor merger between the Milky Way and a small
galaxy can greatly reduce any annihilation signal. For example, if the
Milky Way were to accrete a galaxy with a black hole whose mass is one
tenth of \sgr, then a strong density enhancement near \sgr\ would be
destroyed. 

On the other hand, the more modest density peak produced by
gravitational focusing will persist, even after a merger event,
presuming that dark matter particle orbits in the vicinity of
\sgr\ are randomized. Gravitational focusing will also occur around
other, smaller compact objects in Galaxy.  Consider, for example, a
neutron star moving with considerable velocity through a thermal bath
of dark matter. Figure~3 shows the density enhancement along the
star's direction of motion, and Figure~4 illustrates the intensity of
annihilation radiation from material in a plane containing the path of
the star. The signal is strongest behind the star, and weaker in
front, but the angle-averaged signal strength is not greatly different
from the case where the star is at rest in the bulk dark matter frame.

A neutron star is an improbable source of annihilation radiation in
the form of synchrotron photons; the volume where
there is significant gravitational focusing
is too small to be observable in the case of the neutralino.  The
physics is nonetheless remarkable. Neutralino self-annihilation
through a charged pion channel in a $10^{12}$~G magnetic field will
produce a pion synchrotron signal, since the synchrotron cooling rate
is fast ($<10^{-11}$~s) compared with the pion lifetime
($>10^{-8}$~s).  Furthermore, quantum effects \citep{Bar89} will be
important in the case of a magnetar, with field strength in excess of
$10^{14}$~G.  Then, the emitted photons will have energies comparable
to the typical pion energy.

\section{Conclusion}

Gravitational focusing will generally enhance the annihilation signal
around any massive compact object.  However, because of geometrical
dilution, this signal may not be distinguishable from the background
annihilation radiation. Fortunately, local conditions around a compact
object may cause the annihilation signal to appear in wavebands which
are different from the background. For example, magnetic fields tied
to compact objects can generate a distinctive synchrotron
radiation. Of the sources considered here---neutron stars,
stellar-mass black holes, and massive black holes---only the latter
seems capable of producing a detectable signal by gravitational
focusing alone.  Neutron stars, specifically magnetars, may be the
weakest sources, but they still promise interesting physics involving WIMP
decay products.

Theoretical models of WIMP annihilation in the Galactic Center,
bolstered by recent observations with the Fermi Gamma-ray Space
Telescope \citep[e.g.,][]{Abd-etal10}, have the dark matter
concentrated about \sgr\ as a result of uncertain galactic evolution
and black hole growth dynamics. The main point of this work is that
gravitational focusing will generate {\em some} signal, no matter what
the formation history. While distinctive gamma ray signal could be
lost in a uniform bath of WIMPs, synchrotron radiation may retain some
contrast with the background. The trade-off with synchrotron signal is
that the accretion flows that presumably generate the localized
magnetic field may contaminate the signal. Our best hope may be to
find isolated intermediate mass black holes
\citep{BerZenSil05}. Alternatively we might seek an underluminous
active galactic nucleus. Fortunately, our own Galaxy has one.

%\acknowledgements:

%I am grateful to Ed Bertschinger, Paolo Gondolo, David Kieda, and
%Licia Verde for discussion and comments. The parallel supercomputer
%time used for simulations of dark matter flow around compact objects
%was provided by NASA JPL Supercomputing.
% THIS WAS SO LONG AGO! Maybe it would be more confusing than polite?

\newpage

\begin{figure}
\centerline{\includegraphics[width=\figwidth]{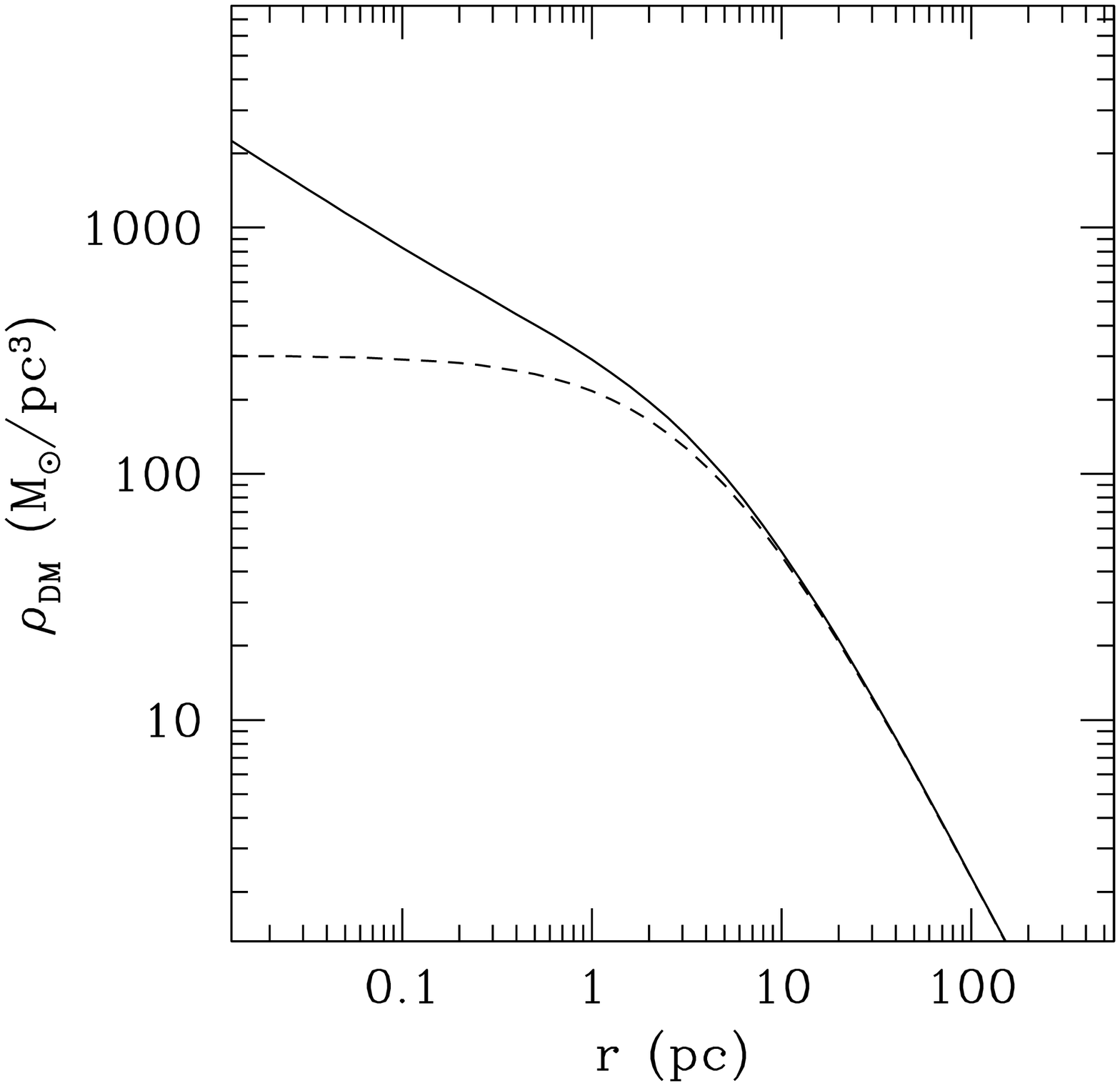}}
 \caption{
   \label{fig1}
   The density enhancement from gravitational focusing of dark matter
   in a density well that approximates the smoothing effect of mergers. 
   The dashed line represents the dark matter without focusing,
   while the solid curve approximates the gravitationally focused profile.
 }
\end{figure}

\newpage

\begin{figure}[b]
\centerline{\includegraphics[width=\figwidth]{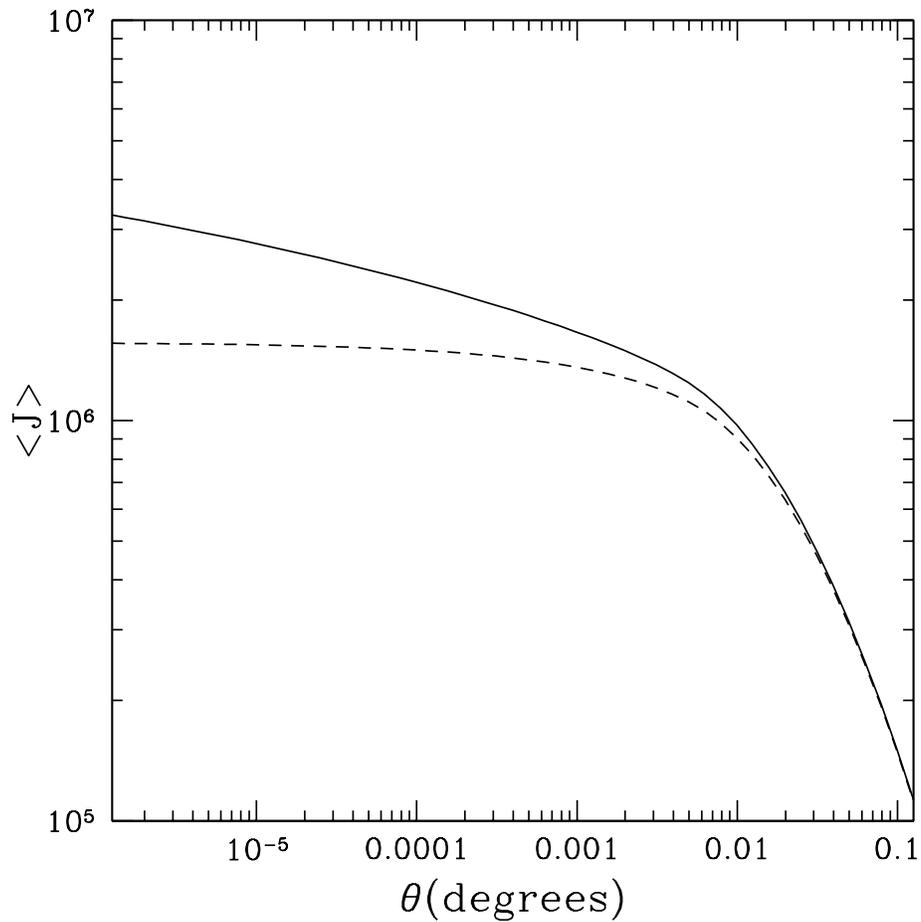}}
\caption{\label{fig2} 
The dimensionless neutralino self-annihilation 
flux $J$ (eq.~3), averaged inside a circular beam
of radius $\theta$, centered on \sgr. As in Fig.~1, the 
dashed line represents a calculation without the effect of gravitational
focusing, while the solid curve includes the focusing effect.}
\end{figure}

\newpage

\begin{figure}
\centerline{\includegraphics[width=\figwidth]{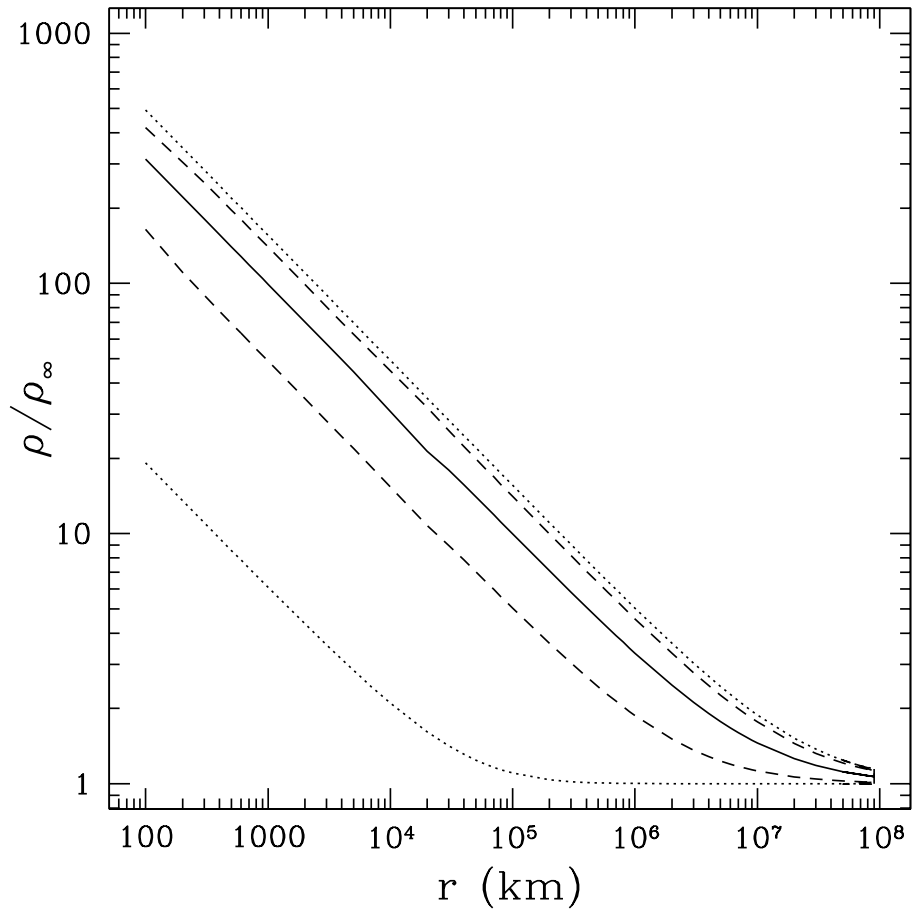}}
\caption{\label{fig3} The density enhancement of thermal dark matter
($\sigma_v = 155$~km/s) in the wake of a neutron star. The solid line
shows the case where the neutron star is at rest with respect to the 
frame of the dark matter bath; the dashed and dotted lines correspond
to neutron star velocities of 200~km/s and 1600~km/s, respectively.
The upper curves show focusing of the downstream flow, lower curves
correspond to upstream density enhancement. In the limit of infinite
star velocity, a downstream density caustic will emerge~\citep{SikWic02}. 
}
\end{figure}

\newpage

\begin{figure}
\centerline{\includegraphics[width=6.5in]{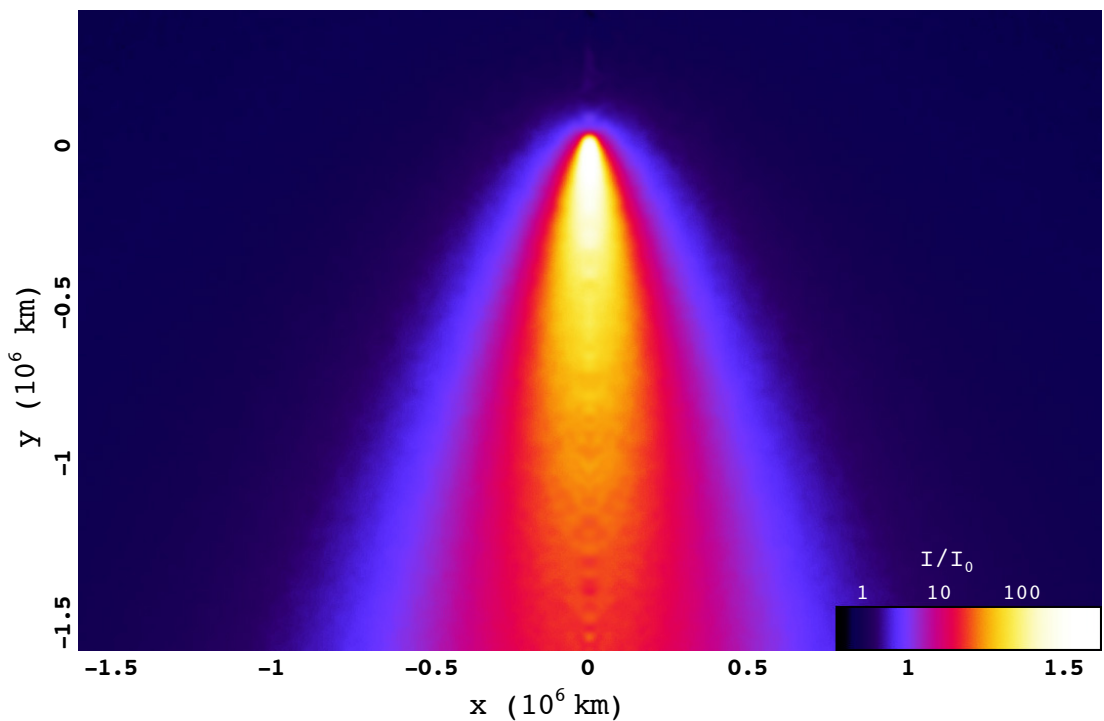}}
%\vspace{11pt}
\caption{\label{fig4} The intensity of self-annihilation radiation a
  plane containing the path of a high-velocity neutron star.  In this
  instance, the neutron star (located at the origin) is moving at
  1000~km/s in a thermal bath of dark matter ($\sigma_v = 200$~km/s).
  The color scale (see inset) corresponds to the log of the intensity
  $I$ (relative to the background signal $I_0$) from dark matter in a
  thin sheet containing this plane, as approxmated from a direct
  simulation of $\sim 10^9$ test particles passing by the star.  }
\end{figure}

\end{document}